\documentclass[preprint,prl,showpacs,amsmath,bibnotes,superscriptaddress,epsfig]{revtex4}
\usepackage{graphicx}

\newcommand\pictc[5]{\begin{figure}
                       \centerline{
                       \includegraphics[width=0.75\columnwidth]{#3}}
                   \protect\caption{\protect\label{fig:#4} #5}
                    \end{figure}            }
\newcommand\pict[4][1]{\pictc{#1}{!tb}{#2}{#3}{#4}}

\newcounter{Fig}

\begin{document}
\begin{sloppy}

\title{Giant Lamb shift in photonic crystals}

\author{Xue-Hua Wang$^{\dag}$}
\author{Yuri S. Kivshar}

\affiliation{Nonlinear Physics Group and Center for Ultra-high
bandwidth Devices for Optical Systems (CUDOS), Research School of
Physical Sciences and Engineering, Australian National University,
Canberra, ACT 0200, Australia}
\homepage{www.rsphysse.anu.edu.au/nonlinear}

\author{Ben--Yuan Gu}
\affiliation{Institute of Physics, Chinese Academy of Sciences,
P.O. Box 603, Beijing 100080, China }

\begin{abstract}
We obtain a general result for the Lamb shift of excited states of
multi-level atoms in inhomogeneous electromagnetic structures and
apply it to study atomic hydrogen in inverse-opal photonic
crystals. We find that the photonic-crystal environment can lead
to very large values of the Lamb shift, as compared to the case of
vacuum. We also predict that the position-dependent Lamb shift
should extend from a single level to a mini-band for an assemble
of atoms with random distribution in space, similar to the
velocity-dependent Doppler effect in atomic/molecular gases.
\end{abstract}

\pacs{42.70.Qs, 32.80.-t, 42.50.Ct.}
 \maketitle
Since the pioneering experiment performed by Lamb and
Retherford~\cite{Lamb} in 1947 and the subsequent theoretical
analysis developed by Bethe~\cite{Bethe}, the study of the Lamb
shift plays an unique role in quantum electrodynamics (QED)
because it provides an excellent test of the QED theory by
comparing its predictions with experimental
observations~\cite{van, Karsh}. Recently, many efforts have been
devoted to the study of various physical effects associated with
the Lamb shift \cite{Vlad,Yu,Chai}.

Photonic crystals (PCs) are a new type of optical material with a
periodic dielectric structure~\cite{s_john}. They can pronouncedly
modify the photonic density of state (DOS) and local DOS leading
to novel quantum-optics phenomena~\cite{sakoda} such as
inhibition~\cite{ES} and coherent control~\cite{Quang} of
spontaneous emission, enhanced quantum interference
effects~\cite{Zhu1}, non--Markovian effects~\cite{Bay,John1}, wide
lifetime distribution~\cite{wangxh1}, non-classic
decay~\cite{wangxh2}, and slop discontinuities in the power
spectra \cite{alva}, etc.

Strong suppression or enhancement of light emission by the PC
environment is expected to modify the Lamb shift. However, very
different predictions for the Lamb shift can be found in
literature. The isotropic dispersion model~\cite{John2} predicts
an anomalous Lamb shift and level splitting for multi-level atoms.
For two-level atoms, the anisotropic model~\cite{Zhu2} suggests
that the Lamb shift should be much smaller than that in vacuum,
while the pseudogap model \cite{Vats} predicts a change of the
Lamb shift of the order of $15\%$ compared to its vacuum value. At
last, a direct extension of the Lamb shift formulism for
multi-level atoms in vacuum to the case of PCs suggests that the
Lamb shift differs negligibly from its vacuum value~\cite{Li}.

Motivated by previous controversial results, in this Letter we
employ the Green's function formalism of the evolution operator to
obtain a general result for the Lamb shift in PCs. We reveal that
in an inhomogeneous electromagnetic environment the dominant
contribution to the Lamb shift comes from emission of real photon,
while the contribution from emission and reabsorption of virtual
photon is negligible, in vast contrast with the case of free space
where the virtual photon processes play a key role. The properties
of the Lamb shift near the band gap are calculated numerically for
an inverse opal PC. We find that the PC structure can lead to {\em
a giant Lamb shift}, and the Lamb shift is sensitive to both the
position of an atom in PCs and the transition frequency of the
related excited level.

We study the Lamb shift in PCs in the framework of nonrelativistic
quantum theory. For an multi-level atom located at the position
${\bf r}$ in a perfect 3D PC without defects, Hamiltonian of the
system can be presented in the form $H=H_0+H_{\rm int}+H_{\rm
ct}$, where the term $H_0$ stands for noninteracting Hamiltonian
and the term $H_{\rm int}$ describes interaction between an atom
and photons, so that
\begin{equation}
H_0+H_{int}= \frac{{\bf p}^2}{2m} + V_a+\hbar\sum_{n\bf{k}}
\omega_{n\bf{k}}a_{n\bf{k}}^{+}a_{n\bf{k}}+\frac{e}{m}{\bf
p}\cdot{{\bf A}({\bf r})} \label{H0}
\end{equation}
with ${\bf A}({\bf
r})=\sum_{n\bf{k}}(\hbar/2\varepsilon_0\omega_{n\bf{k}})^{1/2}[{\bf
E}_n({\bf k}, {\bf r})a_{n{\bf k}}+\mbox{H. C.}]$ being the
quantized vector potential, the second-order term of the vector
potential in Eq.~(\ref{H0}) has been neglected, and $H_{\rm
ct}=(\delta m/m){\bf p}^2/2m$ is a mass-renormalization
counter-term for an electron of observable mass $m$
\cite{John2,Loui}.  The electromagnetic (EM) eigenmodes $\{
\omega_{n{\bf k}},{\bf E}_{n{\bf k}}({\bf r})\}$ in PCs can be
found by the plane-wave expansion method~\cite{Ho}.

We assume that an atom is excited initially, and it stays at the
$l$-th energy level without a photon in the EM field, and denote
$\mid I>=\mid l,0>$ and $\mid F^j_{nk}>=\mid j,1_{n{\bf k}}>$
(i.e. the atom is at the level $j$ and the EM field has a photon
in the state $n{\bf k}$) as the initial and final states of the
system, respectively. The state vector of the system evolves
according to the equation, $\mid \Psi(t)>\equiv U(t)\mid
I>=C_i(t)\mid I>+\sum_{j,n{\bf k}}C^j_{n{\bf k}}(t)\mid
F^j_{nk}>$, with the initial conditions $C_i(0)=1$ and $C_{n{\bf
k}}(0)=0$, where $U(t)$ is the evolution operator. Applying the
Green's function technique to the evolution operator, we obtain
the Fourier transform $C_i(\omega)$ of $C_i(t)$ in the
form~\cite{cohen},
\begin{equation}
C_{i}(\omega)=\frac{1}{2\pi i}\left
[G_{ii}^-(\omega)-G_{ii}^+(\omega)\right ]~,
 \label{Ci}
\end{equation}
with
$G_{ii}^{\pm}(\omega)=\lim_{\eta\rightarrow 0_+}<I\mid
G(z=\omega\pm i\eta)\mid I>,$
where  $G(z)$ is defined by the operator identity
$G(z)(z-H/\hbar)=1$. Projecting this operator identity onto the
one-photon Hilbert space~\cite{lamb} and noting that the
nonvanishing matrix elements of $H_{\rm int}$ are $<F^j_{n{\bf
k}}\mid H_{int}\mid I>$, we obtain the following analytic
expression
\begin{equation}
G^{\pm}_{ii}(\omega)=\lim_{\eta\rightarrow
0^+}\frac{1}{(\omega-\omega_l)-\Delta\pm i[\Gamma/2+\eta]},
 \label{Gii}
\end{equation}
where $\Gamma = \sum_j\alpha_{lj}g({\bf r},\omega-\omega_j)$,
$\Delta=\sum_j(\alpha_{lj}/2\pi)(\omega-\omega_j)\beta({\bf
r},\omega-\omega_j) $, and
\begin{equation}
g({\bf r},\omega)= \displaystyle{\frac{
c^3V_{pc}}{8\pi\omega}\sum_n\int_{BZ}d{\bf k}\mid {\bf E}_{n\bf
k}({\bf
 r})\mid^2\delta(\omega-\omega_{n{\bf
 k}})},\label{Gamma}
\end{equation}

\begin{equation}
 \beta({\bf r},\omega-\omega_j)={\cal P}
{\displaystyle\int^{\omega_{\rm rel}}_0\frac{g({\bf r},\omega ')}
{(\omega-\omega_j-\omega ')\omega '}d\omega '}.
 \label{beta}
\end{equation}
Here $V_{pc}$ is the PC volume, $\omega_{\rm rel}=mc^2/\hbar$ is
the relativistic limit of the photon energy~\cite{Bethe},
$\alpha_{lj}=e^2|{\bf p}_{lj}|^2/3\pi m^2\epsilon_0\hbar c^3$ is
the relative line width of the atomic radiation from the $l$-state
to $j$-state in vacuum, and $\cal P$ stands for the principal
value of the integral. In Eqs.~(\ref{Gamma}) and (\ref{beta}), we
have considered a random orientation of ${\bf p}_{lj}$ and include
the mass-renormalization contribution, respectively
\cite{Loui,John2,Zhu2,Vats,Li}.  The function $g({\bf r},\omega)$
is the local spectral response function (LSRF) proportional to the
photon local DOS.

Equations (\ref{Ci}) and (\ref{Gii}) show that the radiative
correction to the bound level $l$  is determined by the expression
 \begin{equation}
(\omega-\omega_l)=\sum_j\frac{\alpha_{lj}}{2\pi}(\omega-\omega_j)\beta({\bf
r},\omega-\omega_j). \label{shift}
\end{equation}
In the two dispersion models, $|{\bf E}_{n{\bf k}}({\bf
r})|^2=1/V_{pc}$(V$^2$/m$^2$), then Eq. (\ref{shift}) just gives
the results described by Eq.~(6a) of Ref.~\cite{John2} provided we
take $l=1$. For a two-level atom with $j=0,1$, we note that
$\alpha_{11}=0$ due to ${\bf p}_{11}\equiv 0$ (${\bf
p}_{lj}=i(\omega_l-\omega_j)m{\bf r}_{lj}$), and Eq. (\ref{shift})
can be simplified to Eq. (4.9) of Ref. \cite{Vats} by setting
$l=1$ and $\omega_0$ as zero point of energy. In vacuum, $g({\bf
r},\omega')=\omega'$, and by setting $\omega=\omega_l$ in the
right-hand side of Eq. (\ref{shift}), we obtain
\begin{equation}
\Delta^0_l=\frac{e^2}{6\pi^2 m^2\epsilon_0\hbar
c^3}\sum_j-\omega_{jl}|{\bf p}_{lj}|^2\beta({\bf r},-\omega_{jl}),
\end{equation}
where $\omega_{jl}=\omega_j-\omega_l$ and
$\beta({\bf
r},-\omega_{jl})=-\ln(\omega_{rel}/\omega_{jl}+1)\approx
-\ln(\omega_{rel}/\omega_{jl})$.  Because $\beta({\bf
r},-\omega_{jl})$ is a slowly varying function of $\omega_{jl}$,
it is reasonable to make the approximation,
$\omega_{jl}\approx\bar{\omega}-\omega_l$, for $\beta({\bf
r},-\omega_{jl})$ (see also Ref.~\cite{Loui}), with
$\bar{\omega}\gg\omega_l$ being a weighted average of
$\{\omega_j\}$. This approach implies that the dominant
contributions to the Lamb shift come from the emission and
reabsorption of virtual photons (corresponding to the transition
processes from the $l$ level to higher levels), rather than
emission of real photon (corresponding to transition processes
from the $l$ level to lower levels). Noticing that
$\sum_j\omega_{jl}|{\bf p}_{lj}|^2=\hbar
e^2|\psi_l(0)|^2/2\varepsilon_0$, where $\psi_l(0)$ is the wave
function value at the center of an atom in the state $\mid l>$, we
finally obtain a standard nonrelativistic result,
\begin{equation}
\Delta^0_l=\frac{e^4|\psi_l(0)|^2}{12\pi^2m^2\varepsilon_0^2c^3}\ln\frac{\omega_{rel}}{\bar{(\omega}-\omega_l)}.
\label{vacuum}
\end{equation}
Thus, Eq. (\ref{shift}) gives {\em a general result} for a
nonrelativistic radiative correction to a bound level of a
multi-level atom in an inhomogeneous EM system.

We solve  Eq. (\ref{shift}) numerically for an actual PC
structure. For calculating the function $g({\bf r},\omega')$, we
employ an efficient numerical method recently developed in
Ref.~\cite{wangxhr}. For calculating $\beta({\bf
r},\omega-\omega_j)$, we make a reasonable approximation following
the Refs. \cite{Li} and \cite{Bykov}: the dispersion function
$g({\bf r},\omega')$ of a PC vanishes jump-wise at a certain
higher optical frequency $\omega_{\rm op}$, i.e., for
$\omega'>\omega_{\rm op}$, and the PC medium is approximately
treated as a free space with $\varepsilon({\bf r})=1$. We choose
$\omega_{\rm op}$ in such a way that our results are verified to
be  insensitive to perturbations.  $\omega_{\rm op}a/2\pi c\simeq
3.5$ is chosen in our calculations. Furthermore, we distinguish
two different types of integrals for $\beta({\bf
r},\omega-\omega_j)$: the principal integral, when the integrand
in Eq. (\ref{beta}) has a singularity, and the normal integral,
otherwise. With this in hand, we find that the terms for $j<l$ and
for $j>l$ in the right hand side of Eq.~(\ref{shift}) contribute
the principal and normal integrals near $\omega_l$, respectively.
In order to  show this clearly, we assume that
$\omega=\delta+\omega_l$ is a solution of Eq.~(\ref{shift}), and
$|\delta|\ll\omega_{l+1}-\omega_l$, where $\omega_{l+1}$ is the
closest to and higher than the frequency of the level $l$. For
$j<l$, the integrand has a singularity due to
$\delta+\omega_l-\omega_j\geq 0$. But for $j>l$, the integrand has
no singularity due to $\delta+\omega_l-\omega_j<0$.

In PCs, the LSRF $g({\bf r},\omega')$ displays dramatic
fluctuations when the frequency $\omega'$ varies for a given
position. As an example, we demonstrate this in Fig.~1 for an 3D
inverse-opal PC~\cite{Wij} without stacking faults
\cite{Yannopapas}. Thus, the principal integral $\beta({\bf
r},\omega)$ $(\omega>0)$ should be very sensitive to the value of
$\omega$, and the contribution to the integral  comes mainly from
the region near the frequency $\omega$. Figure 2 shows that
$\beta({\bf r},\omega)$ is an oscillatory function of $\omega$.
However, for the normal integral $\beta({\bf r},-\omega)$
$(\omega>0)$,  the fluctuation in $g({\bf r},\omega')$ are
smoothed out after integration, and $\beta({\bf r},-\omega)$ is a
slowly varying function of $\omega$, similar to the case of
vacuum. In Fig. 3, we find a confirmation of this behavior of the
function $\beta({\bf r},-\omega)$. Furthermore, it can be seen
that in a PC the function $\beta({\bf r},-\omega)$ tends to the
limit value of that in vacuum as the frequency $\omega$ grows.
Therefore, the terms with $j>l$ in the right-hand side of Eq.
(\ref{shift}) can be treated similar to the case of vacuum. If we
consider $\bar{\omega}-\omega_l \gg 1$, then the PCs do not bring
about appreciable changes in those terms with $j>l$ compared to
the case of vacuum. Therefore, Eq. (\ref{shift}) can be
approximated as follows
\begin{equation}
\omega-\omega_l-\Delta_l^0\simeq
\sum_{j<l}\frac{\alpha_{lj}(\omega-\omega_j)}{2\pi}{\cal P}
\int^{\omega_{\rm op}}_0\frac{g({\bf r},\omega ')-\omega'}
{(\omega-\omega_j-\omega ')\omega '}d\omega '.
\label{lf}
\end{equation}
Equation (\ref{lf}) shows that, compared to the case of vacuum,
inhomogeneous EM systems lead to an additional contribution to the
Lamb shift that comes mainly from the real photon processes,
rather than the virtual photon precesses, in contrast to the case
of vacuum.

We now apply our result (\ref{lf}) to study the Lamb shift for a
hydrogen atom in the inverse-opal PC. First, we obtain an
interesting result that the PCs environment has no effect on the
$2s$ state due to $\alpha_{2s1s}=0$; this result coincides with
the prediction obtained earlier from the isotropic dispersion
model~\cite{John2}. However, for the $2p$ state, we have
$\Delta^0_{2p}=0$ and $\alpha_{2p1s}\simeq 4\times 10^{-7}$.
Numerical results for the Lamb shift of the $2p$ state are
presented in Fig. 4. We find no level splitting, which differs our
finding from the prediction of the isotropic model~\cite{John2}.
In addition, the Lamb shift depends strongly on not only the
transition frequency but also on the atomic space position,
different from dispersion models \cite{John2,Zhu2,Vats}. The
similar properties can also be found for the $3s$, $3p$, and $3d$
states.

Analyzing the results presented in Fig.~4, we notice that the Lamb
shift can take very large positive or negative values and,
therefore, it can be termed as {\em a giant Lamb shift}. Comparing
the results for the PC with those for vacuum, we find that the
Lamb shift may be enhanced in the PC {\em by one or two orders of
magnitude}. Furthermore, it is significant to point out that the
giant Lamb shift may occur for the transition frequency being
either near or far away from the photonic band gap. The
above-mentioned results are in contrast to the predictions based
on dramatically simplified models~\cite{John2,Zhu2,Vats,Li}. In
Ref. \cite{Li}, a PBG structure was simply treated as an averaged
homogenous medium. This smooths out the contribution to the Lamb
shift from real photon processes that play a key role in
inhomogeneous systems. In the isotropic model \cite{John2},
$g({\bf r},\omega)\sim (\omega-\omega_c)^{-1/2}/\omega$, that
gives an infinite interaction between atom and photons at the band
edge $\omega=\omega_c$ leading to the level splitting and
anomalous Lamb shift. In the anisotropic model \cite{Zhu2},
$g({\bf r},\omega)\sim (\omega-\omega_c)^{1/2}/\omega$, that leads
to coupling interaction near the band edge being smaller than that
in vacuum where $g({\bf r},\omega)=\omega$; it predicts much
smaller Lamb shift than that in vacuum. In pseudogap model
\cite{Vats}, $g({\bf
r},\omega)\sim\omega\{1-h\exp[(\omega-\omega_c)^2/\sigma^2]\}$,
that gives rise to small values of the Lamb shift near a
pseudogap. Clearly, these models lose the main physical
characteristics of the LSRF $g({\bf r},\omega)$ in realistic PCs
that may result in the giant Lamb shift and other significant
effects.

Based upon the position-dependent Lamb shift, we can suggest a
possible experimental approach for verifying our theoretical
predictions: if an assemble of atoms spreads randomly in PCs, the
atoms at different positions have different values of the Lamb
shift. Then the $l$-state levels of many atoms should form a
$l$-state mini-band, similar to the velocity-dependent Doppler
effect in atomic/molecular gases. This mini-band should be
experimentally observable through the emission spectrum of these
atoms.

In conclusion, we have developed a general formalism for
calculating the Lamb shift for multi-level atoms. It is revealed
that the real photon processes play a key pole in inhomogeneous
dielectric structures. Our numerical results for atomic hydrogen
in a 3D inverse-opal PC show that the Lamb shift may be enhanced
remarkably by the PC environment. We have also predicted the
existence of the Lamb shift mini-band for an assemble of atoms
opening up possible ways for experimental observations. We believe
our results provide a deeper insight into the theory of
spontaneous emission in PCs and many applications such as the
development of thresholdless lasers.

This work was supported by the Australian Research Council. The
authors thank Kurt Busch, Judith Dawes, Martjin de Sterke, Sajeev
John, Ross McPhedran, Sergei Mingaleev, and Kazuaki Sakoda  for
useful discussions and suggestions.

\newpage
\pict{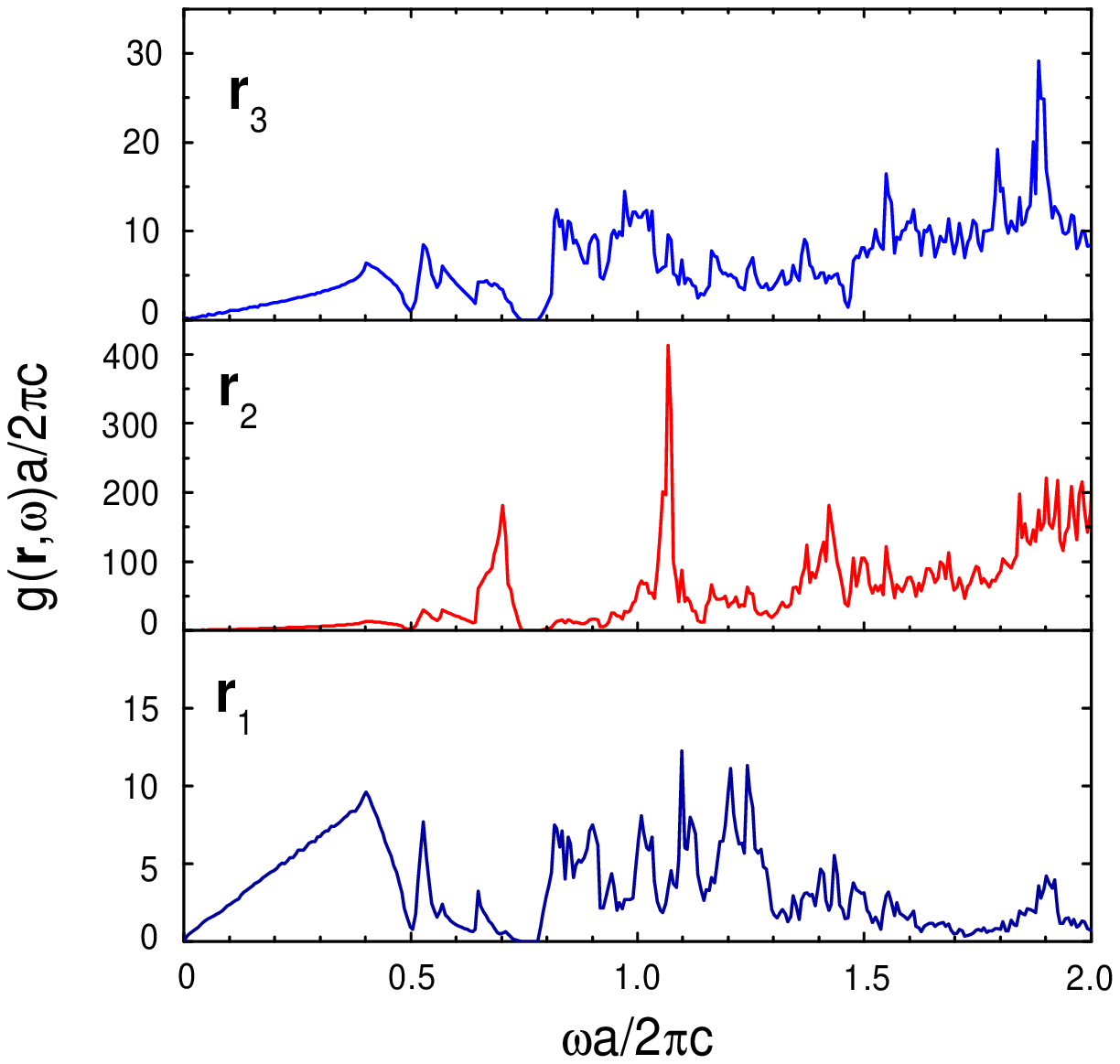}{fig1}{ \small Local spectral response function
$g({\bf r},\omega)$ for an atom placed at three different
positions: ${\bf r}_1=(0,0,0)a$, ${\bf r}_2=(0.34,0,0)a$, and
${\bf r}_3=(0.24,0.24,0)a$ in the inverse-opal photonic crystal
created by air spheres in a medium with $n=3.6$ and $f=0.74$; $a$
is the lattice spacing.}


\pict{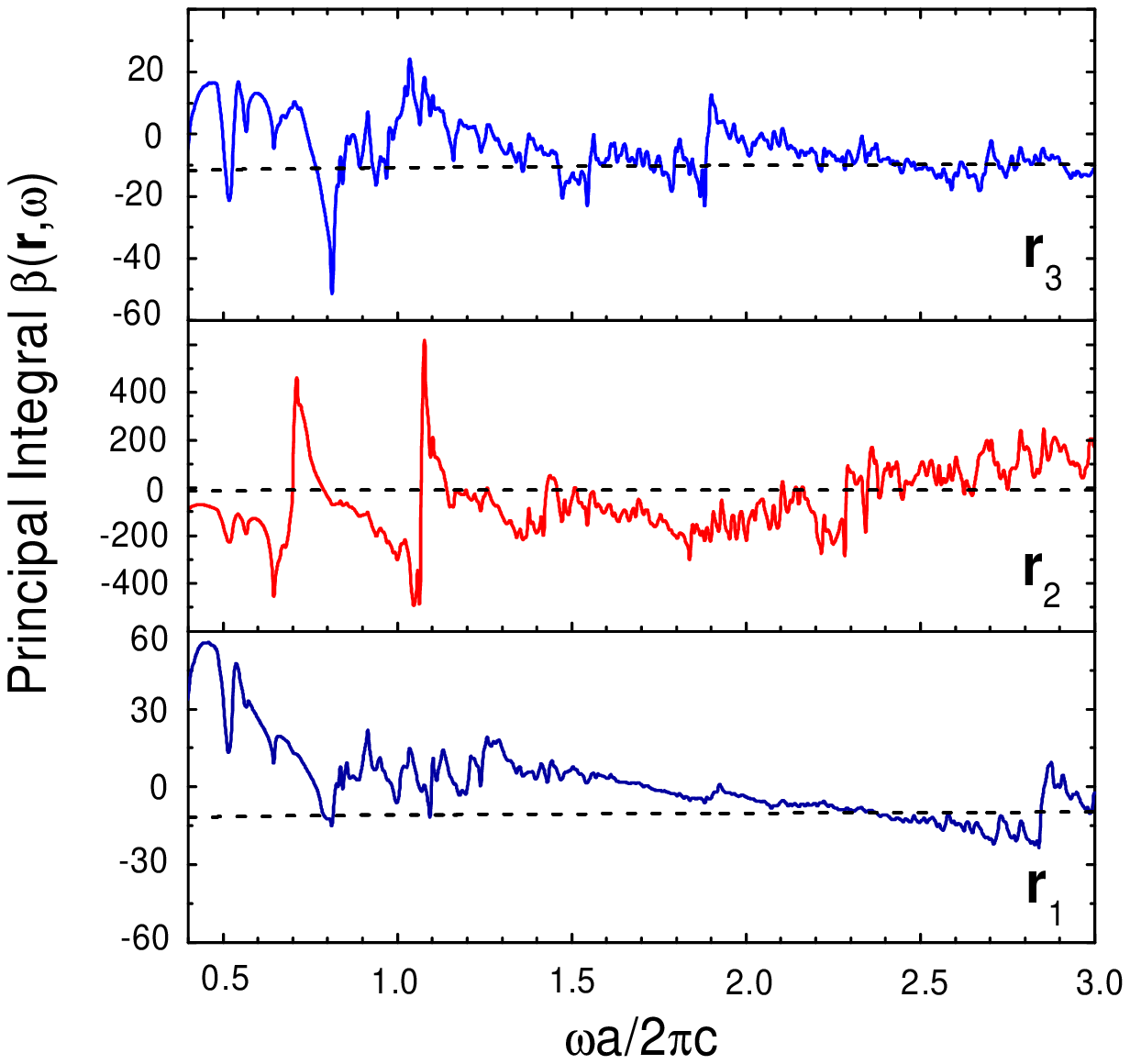}{fig2}{\small Principal integral $\beta({\bf
r},\omega)$ for three different atomic positions in the photonic
crystal. All parameters are the same as in Fig. 1. The dash line
corresponds to the case of vacuum.}

\pict{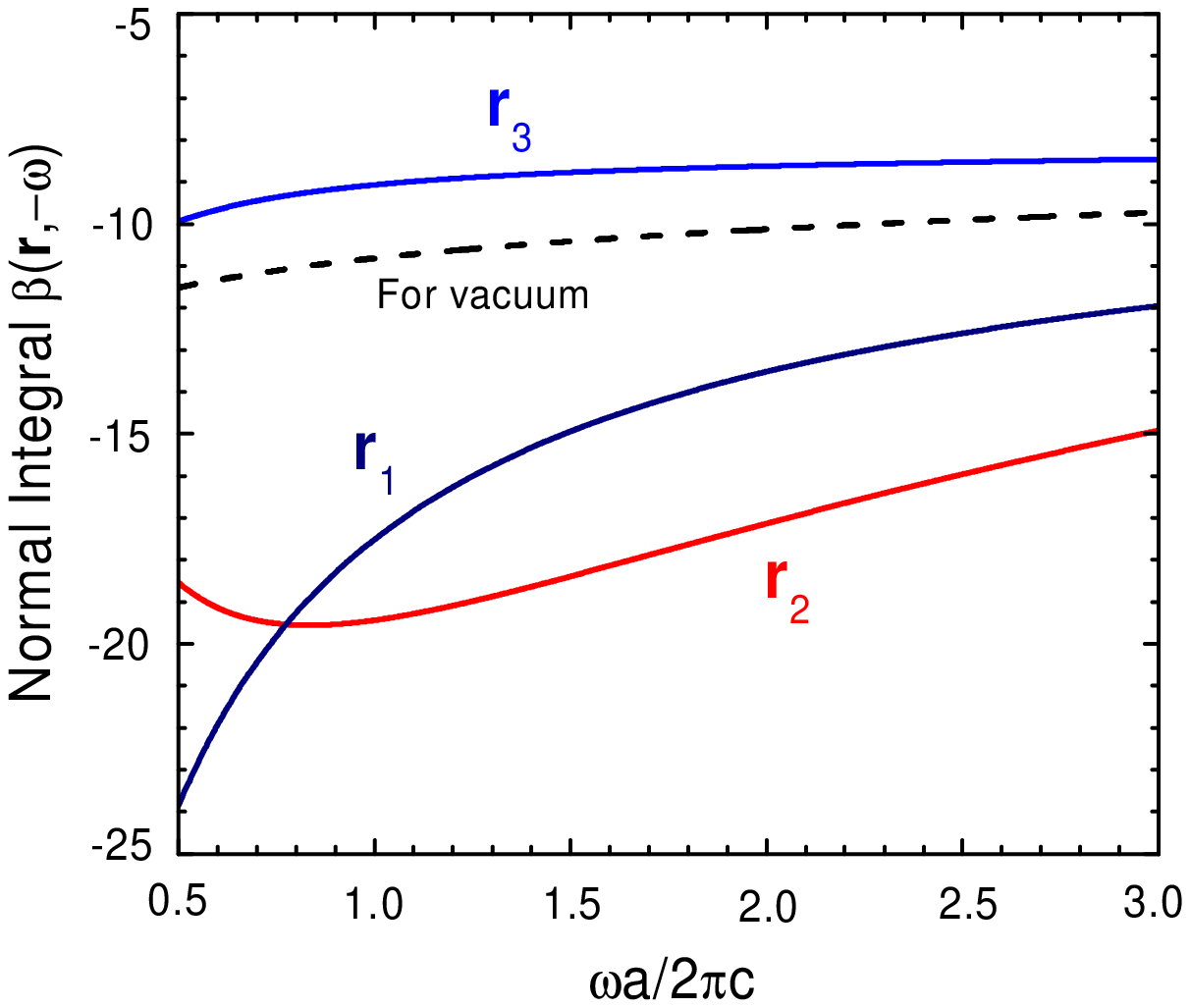}{fig3}{\small Normal integral $\beta({\bf
r},-\omega)$ for three different positions in the photonic
crystal. All parameters are the same as in Fig. 1. The dash line
corresponds to the case of vacuum.}

\pict{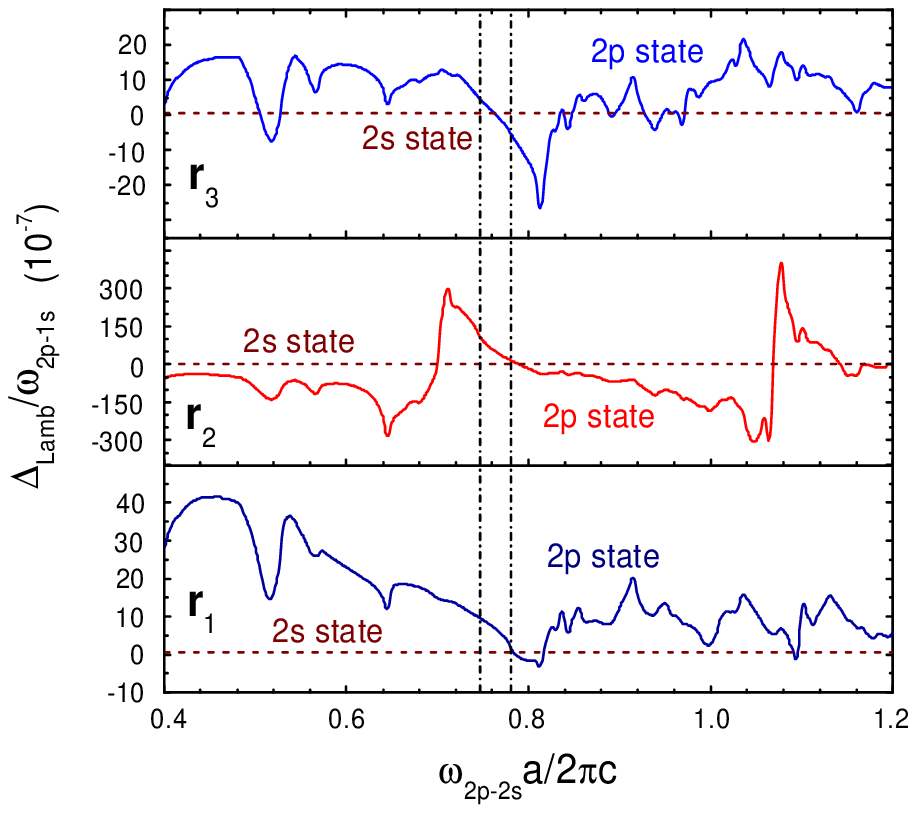}{fig4}{\small The Lamb shift of $2s$ and $2p$ states
as a function of the lattice constant $a$ for atomic hydrogen
located at three different positions in the photonic crystal.  The
Lamb shift of $2s$ state (the dashed lines) is the same as the
value in vacuum. All parameters are the same as in Fig. 1. The
frequency region between two dashed-dot lines is the photonic band
gap.}

\end{sloppy}
\end{document}